\documentclass[twocolumn,aps,amsmath,amssymb,floatfix,superscriptaddress]{revtex4}
\usepackage{graphicx}
\usepackage{dcolumn}
\usepackage{bm}
\usepackage{hyperref}
\usepackage{latexsym}
\usepackage{float}
\usepackage{supertabular}
\usepackage{longtable}

\begin{document}
\title{Promise and Pitfalls of Extending Google's PageRank
Algorithm to Citation Networks}.
\author{S.~Maslov}
\affiliation{Department of Condensed Matter Physics and Materials Science,
Brookhaven National Laboratory, Upton, NY, 11973}
\author{S.~Redner}
\affiliation{Center for Polymer Studies and
and Department of Physics, Boston University, Boston, MA, 02215}

\maketitle

The number of citations is the most commonly-used metric for quantifying the
importance of scientific publications.  However, we all have anecdotal
experiences that citations alone do not characterize the importance of a
publication.  Some of the shortcomings of using citations as a universal
measure of importance include:
\begin{itemize}

\item It ignores the importance of citing papers: a citation from an obscure
  paper is given the same weight as a citation from a ground-breaking and
  highly-cited work.

\item The number of citations is ill-suited to compare the impact of papers
  from different scientific fields.  Due to factors such as size of a field
  and disparate citation practices, the average number of citations per paper
  varies widely between disciplines.  An average paper is cited about 6 times
  in life sciences, 3 times in physics, and $<$1 times in mathematics.
  
\item Many ground-breaking older articles are modestly cited due to a smaller
  scientific community when they were published.  Furthermore, publications
  on significant discoveries often stop accruing citations once their results
  are incorporated into textbooks.  Thus citations consistently underestimate
  the importance of influential old papers.

\end{itemize}

These and related shortcomings of citation numbers are partially obviated by
Google's PageRank algorithm \cite{Google}.  As we shall discuss, PageRank
gives higher weight to publications that are cited by important papers and
also weights citations more highly from papers with few references.  Because
of these attributes, PageRank readily identifies a large number of scientific
``gems'' --- modestly-cited articles that contain ground-breaking results.

In a recent study, we applied \cite{CXMR} Google's PageRank to the citation
network of the premier American Physical Society (APS) family of physics
journals (Physical Review A -- E, Physical Review Letters, Reviews of Modern
Physics, and Physical Review Special Topics).  Our study was based on all
353,268 articles published in APS journals since their inception in 1893
until June 2003 that have at least one citation from within this dataset.
This set of articles has been cited a total of 3,110,839 times by other APS
publications.  Our study is restricted to {\em internal\/} citations ---
citations to APS articles from other APS articles.  Other studies
\cite{google_citation_bolen,eigenfactor} use the PageRank algorithm on a
coarse-grained scale of individual scientific journals to formulate an
alternative to the Impact Factor.

We can think of the set of all APS articles and their citations as a network,
with nodes representing articles and a directed link between two nodes
representing a citation from a {\em citing} article to a {\em cited\/}
article.  In Google's PageRank algorithm \cite{Google}, a random surfer is
initially placed at each node of this network and its position is updated as
follows: (i) with probability $1-d$, a surfer hops to a neighboring node by
following a randomly-selected outgoing link from the current node; (ii) with
probability $d$, a surfer ``gets bored'' and starts a new search from a
randomly selected node in the entire network.  This update is repeated until
the number of surfers at each node reaches a steady value, the Google number.
These nodal Google numbers are then sorted to determine the Google rank of
each node.

The original Brin-Page PageRank algorithm \cite{Google} used the parameter
value $d=0.15$ based on the observation that a typical web surfer follows of
the order of 6 hyperlinks, corresponding to a boredom attrition factor $d=1/6
\simeq 0.15$, before aborting and beginning a new search.  The number of
citation links followed by researchers is generally considerably less than 6.
In \cite{CXMR,CiteRank} we argued that in the context of citations the
appropriate choice is $d=1/2$, corresponding to a citation chain to two links.

\begin{figure}[ht]
 \vspace*{0.cm}
  \includegraphics*[width=0.45\textwidth]{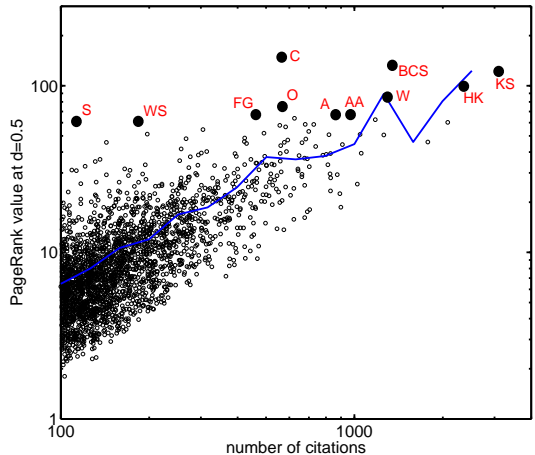}
  \caption{Scatter plot of the Google number versus number of citations for
    APS publications with more than 100 citations.  The top Google-ranked
    papers (filled) are identified by author(s) initials and their details
    are given in Table 1.  The solid curve is the average Google number of
    papers versus number of citations when binned logarithmically.}
\label{extreme}
\end{figure}

The Google number $G$ and number of citations $k$ to a paper are roughly
proportional to each other for $k\agt 50$ \cite{CXMR} so that, on average,
they represent similar measures of importance \cite{FBFM,FFM}.  The outliers
with respect to this proportionality are of particular interest
(Fig.~\ref{extreme}).  While the top APS papers by PageRank are typically
very well-cited, some are modestly-cited and quite old.  The disparity
between PageRank and citation rank arises because the former involves both
the number of citations {\em and\/} quality of the citing articles.  The
PageRank of a paper is enhanced when it is cited by its scientific
``children'' that themselves have high PageRank and when these children have
short reference lists.  That is, children should be influential and the
initial paper should loom as an important ``father figure'' to its children.

The top articles according to Google's PageRank would be recognizable to
almost all physicists, independent of their specialty (Table I).  For
example, Onsager's 1944 exact solution of the two-dimensional Ising model
(the point labeled with $O$ in Fig.~\ref{extreme}) was both a calculational
{\em tour de force} as well as a central development in critical phenomena.
The paper by Feynman and Gell-Mann (FG) introduced the $V-A$ theory of weak
interactions that incorporated parity non-conservation and became the
``standard model'' of the field.  Anderson's paper (A), ``Absence of
Diffusion in Certain Random Lattices'' gave birth to the field of
localization and was cited for the 1977 Nobel prize in physics.  Particularly
intriguing are the articles ``The Theory of Complex Spectra'', by J. C.
Slater (S) and ``On the Constitution of Metallic Sodium'' by E.  Wigner and
F. Seitz (WS) with relatively few APS citations (114 and 184 respectively as
of June 2003).  Slater's paper introduced the determinant form of the
many-body wavefunction that is so ubiquitous that the original work is no
longer cited.  Similarly, the latter paper introduced the Wigner-Seitz
construction that constitutes an essential curriculum component in any
solid-state physics course.

\begin{widetext}
{\small\begin{longtable}{|>{\hfill}p{0.19in}|>{\hfill}p{0.31in}|>{\hfill}p{0.22in}|p{1.25in}|p{2.1in}|p{2.1in}|}

\caption{Top non-review-article APS publications ranked according to PageRank
  (PR, first column), the citation number (CNR, second column), and CiteRank
  \cite{CiteRank} (CR, third column).  The full database of PageRank and
  CiteRank values of APS publications is accessible at \cite{CiteRank_web}.
}\label{top-10}\\

\hline
PR& CNR & CR &{Publication} & Title & Author(s) \\ \hline&&&&& \\[-1em]\hline \endhead
1&54&42&  PRL {\bf 10,} 531 (1963)&            Unitary Symmetry \&  Leptonic Decays & Cabibbo (C)\\ \hline
2&5&10&  PR {\bf 108}, 1175 (1957)&       Theory of  Superconductivity&  Bardeen, Cooper \& Schrieffer (BCS)\\ \hline
3&1&1&  PR {\bf 140}, A1133 (1965)&       Self-Consistent Equations \ldots  &  Kohn \&  Sham (KS)\\ \hline
4&2&2& PR {\bf 136}, B864 (1964)&        Inhomogeneous Electron Gas&   Hohenberg \& Kohn (HK)\\ \hline
5&6&58&  PRL {\bf 19}, 1264 (1967)&         A Model of Leptons&   Weinberg (W)\\ \hline
6&55&37& PR {\bf 65}, 117 (1944)&            Crystal Statistics \ldots&   Onsager (O)\\ \hline
7&95&293&   PR {\bf 109}, 193 (1958)&        Theory of the Fermi Interaction  & Feynman \&  Gell-Mann (FG)\\ \hline
8&17&13& PR {\bf 109}, 1492 (1958)&           Absence of Diffusion in \ldots &   Anderson (A)\\ \hline
9&1853&133& PR {\bf 34}, 1293 (1929)&             The Theory of Complex Spectra  &   Slater (S)\\ \hline
10&12&11& PRL {\bf 42}, 673 (1979)&          Scaling Theory of Localization & Abrahams, Anderson, {\it et al.} (AA)\ \\ \hline
11&712&106& PR {\bf 43}, 804 (1933)&          \ldots  Constitution of Metallic Sodium  & Wigner \& Seitz (WS)\\ \hline
\end{longtable}
}
\end{widetext}

The PageRank algorithm was originally developed to rank webpages that are
connected by hyperlinks and not papers connected by citations. The most
important difference between these two networks is that, unlike hyperlinks,
citations cannot be updated after publication.  The constraint that a paper
may only cite earlier works introduces a time ordering to the citation
network topology.  This ordering makes aging effects much more important in
citation networks than in the world-wide web.  The habits of scientists
looking for relevant scientific literature are also different from web
surfers.  Apart from the already-mentioned shorter depth search ($1/d=2$),
scientists often start literature searches with a paper of interest that they
saw in a recent issue of a journal or heard presented at a conference.  From
this starting point a researcher typically follows chains citations that lead
to progressively older publications.

These observations led us to modify the PageRank algorithm by initially
distributing random surfers exponentially with age, in favor of more recent
publications \cite{CiteRank}.  This algorithm --- CiteRank --- is
characterized by just two parameters: $d$ --- the inverse of the average
citation depth, and $\tau$ --- the time constant of the bias toward more
recent publications at the start of searches.  The optimal values of these
parameters \cite{CiteRank} that best predict the rate at which publications
acquire recent citations are $d=0.5$ and $\tau=2.6$ years for APS
publications. With these values the output of the CiteRank algorithm
quantifies the relevance of a given publication in the context of currently
popular research directions, while that of PageRank corresponds to its
``lifetime achievement award''.  The database with the results of application
of both algorithms to the citation network of APS publications can be
accessed online at \cite{CiteRank_web}.

Google's PageRank algorithm and its modifications hold great promise for
quantifying the impact of scientific publications.  They provide a meaningful
extension to traditionally-used importance measures, such as the number of
citations to articles and the impact factor for journals.  PageRank
implements, in a simple way, the logical notion that citations from more
important publications should contribute more strongly to the rank of a cited
paper.  PageRank also effectively normalizes the impact of papers in
different scientific communities \cite{community}.  Other ways of attributing
a quality for citations would require detailed contextual information about
citation themselves, features that are presently unavailable.  PageRank
implicitly includes context by incorporating the importance of citing
publications.  Thus PageRank represents a computationally simple and
effective way to evaluate the relative importance of publications beyond
simply counting citations.

We conclude with some caveats.  It is very tempting to use citations and
their refinements, such as PageRank, to quantify the importance of
publications and scientists \cite{H}, especially as citation data becomes
increasingly convenient to obtain electronically.  In fact, the $h$-index of
any scientist, a purported single-number measure of the impact of an
individual scientific career, is now easily available from the Web of Science
\cite{web}.  However, we must be vigilant for the overuse and misuse of such
indices.  All the citation measures devised thus far pertain to popularity
rather than to the not-necessarily-coincident attribute of intrinsic
intellectual value.  Even if a way is devised to attach a high-fidelity
quality measure to a citation, there is no substitute for scientific judgment
to assess publications.  We need to avoid falling into the trap of relying on
automatically generated citation statistics for accessing the performance of
individual researchers, departments, and scientific disciplines, and
especially of allowing the evaluation task to be entrusted to administrators
and bureaucrats \cite{IMU_report}.

\acknowledgments{Work by SM at Brookhaven National Laboratory was carried out
  under Contract No. DE-AC02-98CH10886, Division of Material Science,
  U.S. Department of Energy.  SR gratefully acknowledge financial support
  from the US National Science Foundation grant DMR0535503.}


\begin{thebibliography}{99}

\bibitem{Google} S. Brin and L. Page, {\it The anatomy of a large-scale
  hypertextual {Web} search engine}, Computer Networks and ISDN Systems,
  {\bf 30}, 107 (1998).

\bibitem{CXMR} P. Chen, H. Xie, S. Maslov, and S. Redner, {\it Finding
    Scientific Gems with Google}, J. of Informetrics {\bf 1}, 8 (2007).

\bibitem{google_citation_bolen}
J. Bollen, M. A. Rodriguez, and H. Van de Sompel
{\it Journal Status}, Scientometrics {\bf 69}, 669, (2006).

\bibitem{eigenfactor}
C. Bergstrom, {\it Eigenfactor: measuring the value and presitige of scholarly journals.} 
College and Research Libraries News {\bf 68}, 5 (2007). 

\bibitem{CiteRank} D. Walker, H. Xie, K.-K Yan, S. Maslov,
{\it Ranking scientific publications using a model of network traffic.},
J. Stat.\ Mech.\ {\bf 6}, P06010 (2007).

\bibitem{FBFM} S. Fortunato, M. Boguna, A. Flammini, and F. Menczer, {\it How
    to make the top ten: Approximating PageRank from in-degree},
    In Proceeding of WAW2006 (2006). arXiv:cs/0511016.

\bibitem{FFM}  S. Fortunato, A. Flammini, and F. Menczer, {\it Scale-free
    network growth by ranking}, Phys.\ Rev.\ Lett.\ {\bf 96}, 218701 (2006).


\bibitem{CiteRank_web}  http://www.cmth.bnl.gov/$\sim$maslov/citerank

\bibitem{community} H. Xie, K.-K. Yan, and S. Maslov,
{\it Optimal ranking in networks with community structure},
Physica A {\bf 373}, 831 (2007). arXiv:physics/0510107.


\bibitem{H} J. Hirsch, {\it An index to quantify an individual's scientific
    research output}, Proc.\ Natl.\ Acad.\ Sci.\ (USA) {\bf 102}, 16569
  (2005).

\bibitem{web} http://apps.isiknowledge.com/


\bibitem{IMU_report} See also R. Adler, J. Ewing, P.  Taylor, {\it Citation
    Statistics}, Report from the International Mathematical Union (IMU),
  (2008),
  http://www.mathunion.org/fileadmin/IMU/Report \\
  /CitationStatistics.pdf
\end{thebibliography}
\end{document}